\input{aipcheck}

\documentclass[
    ,final            
  ]
  {aipproc}

\layoutstyle{6x9}
\newcommand{\msun}{$M_{\odot}$}

\newcommand{\teff}{$ T\rm_{eff}$}
\newcommand{\logg}{log$g$}

\usepackage{graphicx}

\begin{document}

\title{Anyone out there? Galactic Halo Post-AGB stars}

\classification{97.60}
\keywords      {stars: evolution -- stars: AGB and post-AGB -- Galaxy: halo}

\author{S. Weston}{
  address={University of Hertfordshire, College Lane, Hatfield, Hertfordshire, AL10 9AB, UK}
}

\author{R. Napiwotzki}{
}

\author{S. Catal\'an}{
}

\begin{abstract}
We present results of a survey of post-asymptotic giant branch stars
(post-AGBs) at high galactic latitude. To date, few post-AGB stars are
known throughout the Galaxy and the number of known members of the
older populations like the galactic halo is even smaller. This study
looks at the number of post-AGB stars which are produced using
different synthetic population methods and compare the results with
observations. The resulting synthetic populations are compared to
observational results from a complete and studied subsample from the
photographic Palomar-Green (PG) survey (with high resolution
spectroscopic follow-up for post-AGB candidates) and the SDSS
spectroscopic database. The results show only two candidate post-AGB
stars in a complete subsample of the PG survey spanning 4200 deg$^{2}$
and one in the SDSS database. We discuss and explore any observational
biases which may cause the result. If found to be truely
representative of the halo population, one can expect the majority of
Population II stars to fail to ascend the AGB and evolve through other
evolutionary channels such as the extended horizontal branch.
\end{abstract}

\maketitle


\section{Introduction}
\label{sec:intro}
In the standard scenario, all low and intermediate mass stars should
evolve up the asymptotic giant branch (AGB) and enter the white dwarf
cooling sequence as post-AGB stars. It is known that extreme
horizontal branch (EHB) stars become white dwarfs without an
AGB/post-AGB phase but their numbers are estimated to be
low \cite[$\sim$1\%,][]{dri85,heb86,saf94}. The number of known PNe
and post-AGB stars in the galactic halo is quite small. The Torun
catalogue \cite{szc07} is the most complete compilation of known
post-AGB stars. However, most of the objects included in this
catalogue were detected due to their IR excess including all of the
IRAS objects from \cite{sua06}. This can be expected to introduce a
bias in favour of higher mass pop. I post-AGB stars with dense
envelopes. Only 20 objects in the catalogue are possible or probable
pop. II stars. \\ Higher mass post-AGB stars have suffered from strong
mass loss and are evolving quickly. As a result they can remain
enshrouded in their circumstellar envelope during most of this
phase. Extinction makes the star difficult to detect in the optical,
but they are easily found by the IR surveys mentioned above. Low mass
post-AGB stars (of pop. II) experience relatively small mass loss and
evolve slowly. As a result little or no IR excess is expected, but the
stellar radiation is essentially unabsorbed.\\ Classifying post-AGB
stars is difficult as some types of object have a similar
appearance. Examples include, horizontal branch stars (HB), extreme HB
(EHB), post-EHB (pEHB) and hot, massive MS stars. Accurate parameters
and sometimes only detailed chemical abundance analyses tells them
apart from post-AGB stars. However, the number of stars contaminating
the post-AGB region is expected to be small and our findings can only
increase the number of objects found in the post-AGB region. \\ Here
we describe the results of systematic searches for high galactic
latitude post-AGB stars facilitating optical and UV data. Results are
compared with expectations based on the standard scenarios and we
discuss the implications.

\section{Post-AGB Stars in High Galactic Latitude Surveys}
\begin{figure}
 \includegraphics [scale=0.6] {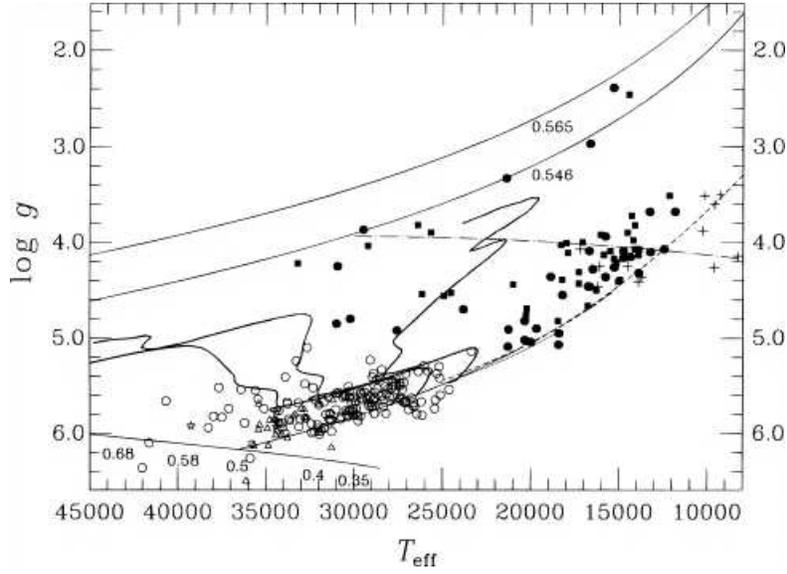}
 \caption{The \teff-\logg\ diagram of the followed-up PG region. This
   table is Fig.\, 5 in Saffer et al. (1997). The solid symbols are halo
   B-type star candidates and the squares are the complete sample
   (details in this and their paper). The evolutionary tracks from low
   to high gravity are firstly, the two solid lines are the post-AGB
   tracks of Sch\"onberner (1983) for the stated masses. The dash-dot line is
   the pop. I ZAMS, the various dashed, dotted and undulating curves
   are HB and post-HB tracks with the solid He ZAMS at the bottom.}
 \label{fig:saffer_hr}
\end{figure}

\textbf{\emph{Sloan Digital Sky Survey (SDSS):}} 
SDSS perfomed an imaging survey of 12,000 square degrees of the
northern sky, mostly with $|b|>10^{\circ}$. An extensive spectroscopic
database of over 1.6 million objects followed-up is available. In a
brute force apporach to identify post-AGB stars in SDSS we carried out
Balmer line fitting of all objects bluer than $g'-r'<0.0$. Quasars and
other extra galactic objects were identified and discarded. \teff\
and \logg\ of the remaining objects were compared with post-AGB
tracks. In the end \textbf{only one} star remained with parameters
compatible with a post-AGB nature. This is obviously at odds with any
plausible estimate of post-AGB numbers. However, one has to have in
mind possible biases in the selection of SDSS targets for
follow-up. Such biases include some objects saturating, photometry not
being unique and post-AGBs are low priority for follow-up. Some
post-AGB objects are known in the SDSS survey area \cite{szc07} but
were not included in the spectroscopic database. This is likely due to
one of the reasons mentioned above.\\
\textbf{\emph{Palomar-Green (PG) Catalogue:}} 
In an effort to double check the completely unexpected outcome of our
search of the SDSS database, we studied findings from the PG
survey. This is a 10,714 square degrees photographic survey of
UV-excess objects of high galactic latitude \cite{gre86}. 1874 objects
were selected for low resolution spectroscopic follow-up based upon
the criteria $U-B<-0.46$ and given a spectral classification. Due to
the low resolution of the spectra the classifications were quite broad
or mistakenly labelled HBB, sd, sdB and sdBO which with higher
resolution follow-up later determined to be post-AGB, HBB, pHBB, pEHB,
and pop. I and II main sequence stars. \citet{saf97} carried out
intermediate resolution follow-up of a collection of stars categorised
as above and the resulting
\teff-\logg\ diagram is shown in Fig.\,\ref{fig:saffer_hr}.  The
interesting objects for our study are the ten post-AGB candidates
which are found near the \cite{sch83} post-AGB tracks. However, also
apparent from the other tracks are the ambiguity of the candidates
with the hotter objects possible post-EHBs and other potential pop. I
MS stars. Therefore, to confidently classify the different
types, high resolution spectroscopy is required to determine chemical
compositon. In an effort to contain a complete sample for their study
and priortise objects for follow-up observations, \cite{saf97}
selected three fields (see Fig.\,\ref{fig:plots_0546}) and a
brightness limit $B_{\rm{PG}}<14.7$. Only three of the ten post-AGB
candidates fulfill the criteria, PG1212+369, PG1243+275 and
PG2120+062. The high resolution spectroscopy revealed that PG1212+369
had a close secondary component which it had probably interacted with,
ruling it out as a post-AGB candidate \citep{ham97}. PG1243+275 is
metal-poor and its abundance makes it a strong post-AGB
candidate. PG2120+062 was confirmed as a post-AGB star through its CNO
depletion. Further details of the object can be found
in \cite{lyn04}. Thus only two post-AGBs observed within the 4200
deg$^{2}$ region of sky followed-up for the complete sample. The
position of all the candidates in the \teff-\logg\ diagram suggests
that they are low mass ($M < 0.55$\msun). This is in agreement with
observed mass distributions for similar WD
populations \citep{pau06,lie05}.

\section{Synthetic Population Predictions}
\label{sec:synth_pred}
We generate a synthetic post-AGB galactic population using an
adaptation of the WD Monte Carlo simulation of \cite{nap09}. The
simulation uses the galactic model structure of \cite{rob03} to
randomly assign locations of a large number of stars based on observed
densities. Depending on population membership each star is given a
metallicity, an initial mass and kinematical properties. Number
densities of the three populations (thin disc, thick disc and halo)
are calibrated with the local population based on the Supernova Type
Ia Survey \cite[SPY,][]{pau06,nap03}. The evolution of each individual
star up to the post-AGB phase is determined from the Padova
evolutionary tracks \cite{gir00} and refs. therein. The tracks give a
metallicity range of Z=0.0001--0.1 corresponding to a [Fe/H] of $-$2.3
to 0.95. This fully covers the range from observed populations which
are reproduced in the simulation. All stars which are not old enough
to have evolved to the tip of the AGB are discarded. The evolution of
the remaining stars is followed through the post-AGB phase and the WD
cooling sequence. We performed simulations using various mass and
metallicity post-AGB tracks
of \cite{sch83}, \cite{vas93}, \cite{vas94}, \cite{blo95} and
\cite{wei09} to compute \teff\ and \logg. The masses range from 
0.524--0.943\msun\ and metallicities $Z$=0.0005 to 0.04 (equivalent to
[Fe/H]=$-$1.6 to 0.3). All post-AGB tracks used for our analysis see
the star leave the AGB as hydrogen-burners. He-burners evolve slower
and thus would produce an even higher number of observable stars. The
higher mass post-AGB stars evolve much quicker to hotter temperatures
and so will spend less time on the top of the H-R diagram and
resulting in less post-AGB stars at a given time. The final post-AGB
population is normalised to the local WD population as described
in \cite{nap09}. \\
\textbf{Simulation of the Complete PG Subsample:} We simulated the 
\cite{saf97} sample of post-AGB
candidates by selecting the stars from the same fields applying the
same brightness limit ($B_{\rm{PG}}<14.7$) as those for the complete
sample defined in \cite{saf97}. We applied a temperature criterion of
14,000--34,000K. This was defined at the low end due to the PG $U-B$
cutoff criteria and the photometric uncertainty attached to this, and
the top end by the hottest found post-AGB candidate in their
sample. These criteria are conservative and can be interpreted as a
lower limit on the number of stars which should be observed in that
survey.
\section{The Results}
\begin{figure}[t]
\begin{minipage}{0.5\textwidth}
 \includegraphics [width=\textwidth, height=0.25\textheight] {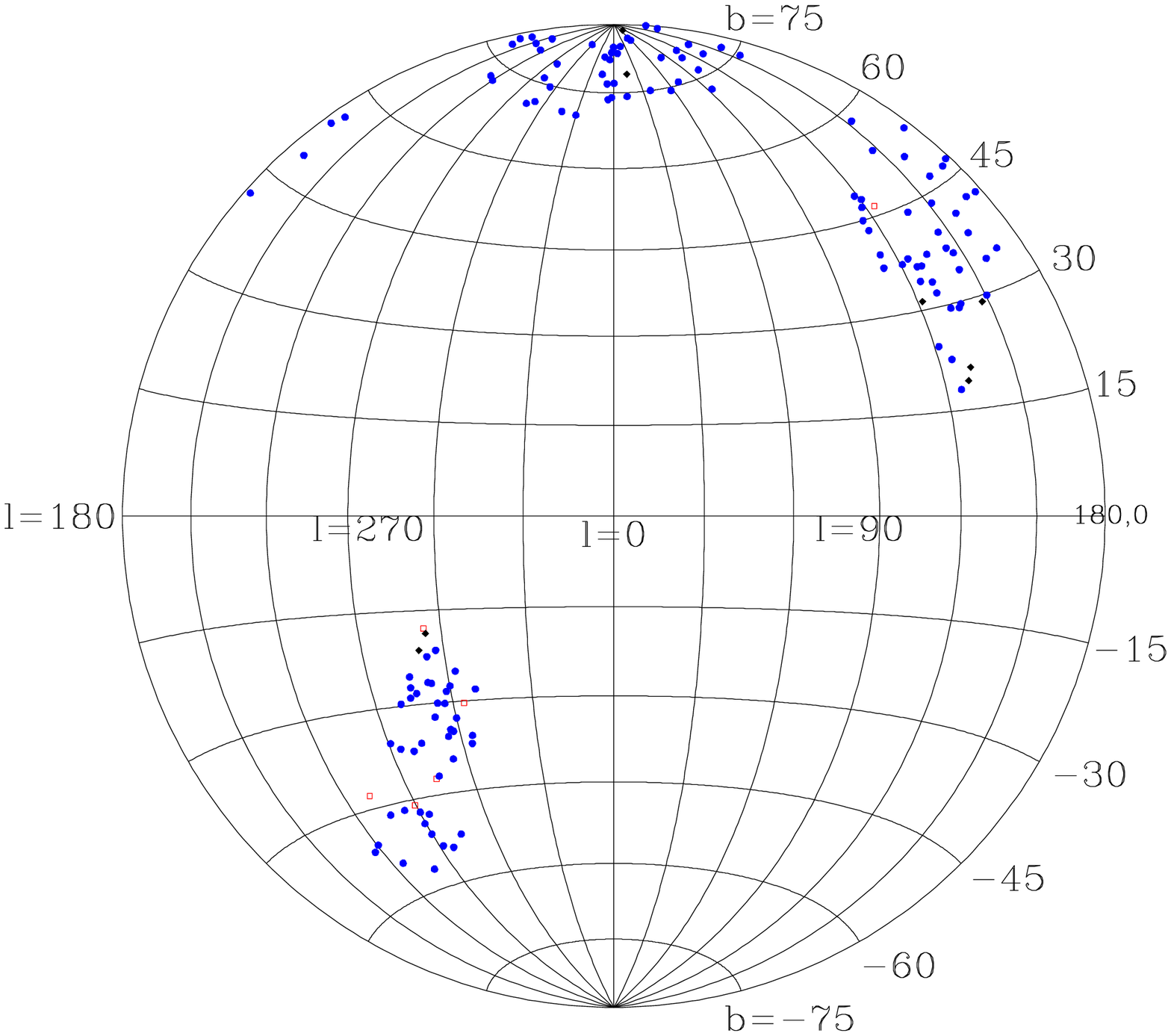}  
\end{minipage}
\begin{minipage}{0.5\textwidth}

 \centering
 \includegraphics [width=\textwidth, height=0.25\textheight] {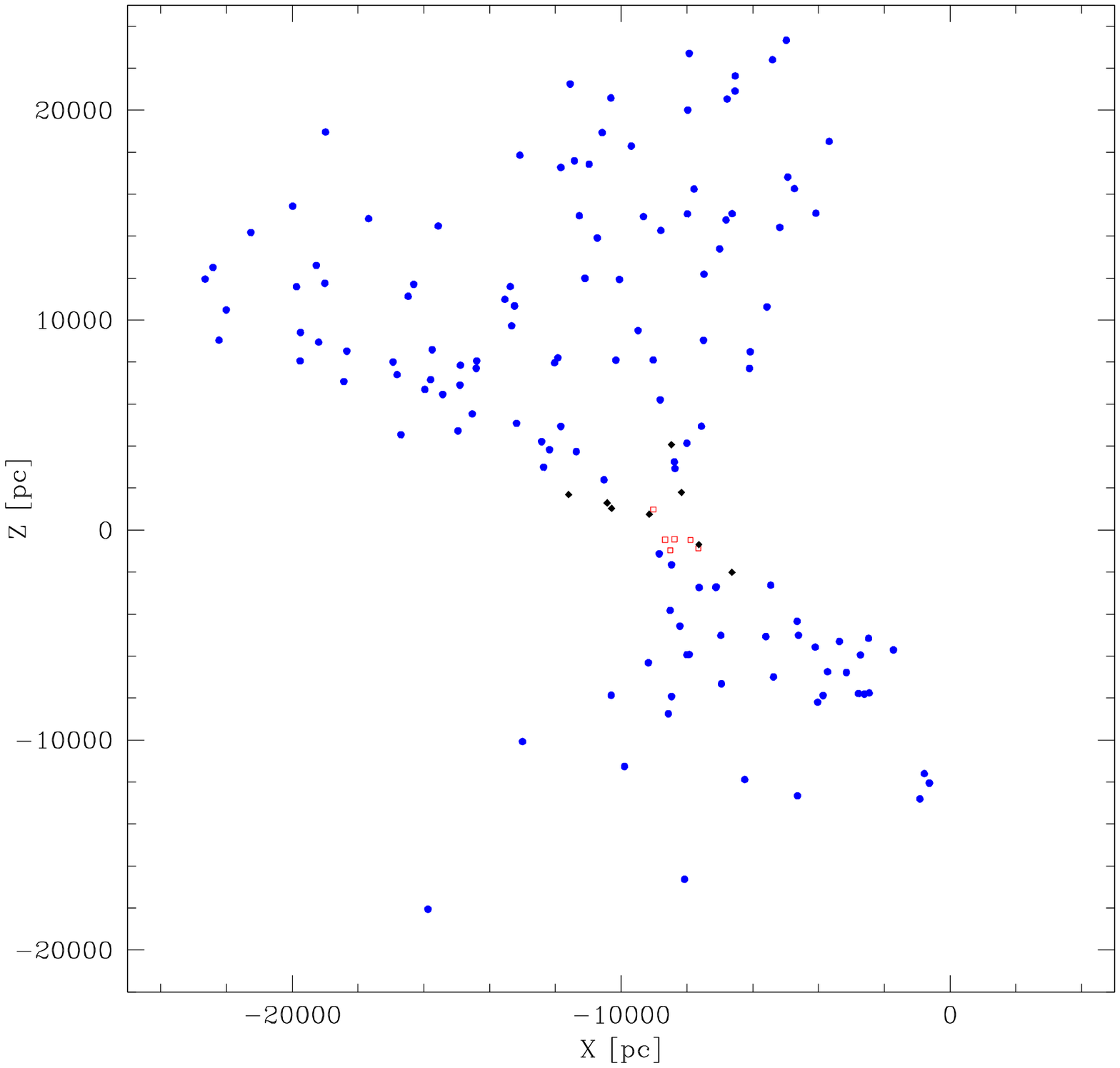}
\end{minipage}
\caption{\emph{Left:} An Aitoff-Hammer projection of the
 post-AGB population from the three selected complete regions in
 galactic coordinates. This example is for a 0.546\msun\ assumed
 post-AGB mass. The red, square, open symbols represent thin disc
 post-AGBs, the black, filled, diamonds the thick disc and the blue
 filled circles the halo. \emph{Right:} A spatial coordinate
 projection of the post-AGB population. Note the galactic centre
 is at vector [X,Z]=(0,0) and the objects converge to our Sun's
 position in the galaxy at approximately
 [X,Z]=(8500,0).}  
 \label{fig:plots_0546}
\end{figure}

\begin{table}
\centering
\begin{minipage}{\textwidth}
\begin{tabular}{lllccccc}
    \hline
    \tablehead{1}{l}{b} {Post-AGB Mass\\\msun}
    & \tablehead{1}{l}{b} {MS Mass\\\msun}
    & \tablehead{1}{l}{b} {MS Met.\\(Z)}  
    & \tablehead{1}{c}{b} {N$^{o}$ \\thin disc}
    & \tablehead{1}{c}{b} {N$^{o}$ \\thick disc} 
    & \tablehead{1}{c}{b} {N$^{o}$ \\halo}
    & \tablehead{1}{c}{b} {Total}
    & \tablehead{1}{c}{b} {Grid \\Ref.}\\ 
    \hline
0.524 & 1.00 & 0.021 & $ 11 \pm 2 $ & $ 36 \pm 4 $ & $ 160 \pm 9 $ & $ 208 \pm 10 $ &\cite{blo95}\\
0.546 & 0.80 & 0.021 & $ 13 \pm 3 $ & $ 28 \pm 4 $ & $ 59 \pm 5 $ & $ 99 \pm 8 $ &\cite{sch83}\\
0.565 & 1.00 & 0.021 & $ 0 \pm 1 $ & $ 1 \pm 1 $ & $ 13 \pm 3 $ & $ 15 \pm 3 $ &\cite{sch83}\\
0.605 & 3.00 & 0.021 & $ 0 \pm 0 $ & $ 0 \pm 1 $ & $ 16 \pm 3 $ & $ 16 \pm 3 $ &\cite{blo95}\\
0.569 & 1.00 & 0.016 & $ 1 \pm 1 $ & $ 3 \pm 1 $ & $ 97 \pm 7 $ & $ 100 \pm 7 $ &\cite{vas93} \\
0.597 & 1.50 & 0.016 & $ 1 \pm 1 $ & $ 1 \pm 1 $ & $ 32 \pm 4 $ & $ 34 \pm 4 $ &\cite{vas93} \\
0.633 & 2.00 & 0.016 & $ 0 \pm 0 $ & $ 0 \pm 0 $ & $ 9 \pm 2 $ & $ 9 \pm 2 $ &\cite{vas93} \\
0.530 & 1.20 & 0.008 & $ 1 \pm 1 $ & $ 1 \pm 1 $ & $ 34 \pm 4 $ & $ 36 \pm 4 $ &\cite{wei09} \\
0.531 & 1.00 & 0.004$^{\alpha}$ & $ 0 \pm 1 $ & $ 2 \pm 1 $ & $ 37 \pm 4 $ & $ 40 \pm 4 $ &\cite{wei09} \\
0.533 & 1.20 & 0.004 & $ 1 \pm 1 $ & $ 2 \pm 1 $ & $ 35 \pm 4 $ & $ 37 \pm 4 $ &\cite{wei09} \\
0.623 & 1.50 & 0.001 & $ 0 \pm 0 $ & $ 2 \pm 1 $ & $ 35 \pm 4 $ & $ 37 \pm 4 $ &\cite{vas93} \\
0.663 & 2.00 & 0.001 & $ 0 \pm 0 $ & $ 0 \pm 0 $ & $ 6 \pm 2 $ & $ 7 \pm 2 $ &\cite{vas93} \\
0.534 & 1.00 & 0.0005$^{\alpha}$ & $ 1 \pm 1 $ & $ 2 \pm 1 $ & $ 36 \pm 4 $ & $ 39 \pm 4 $ &\cite{wei09} \\
0.599 & 2.00 & 0.0005$^{\alpha}$ & $ 0 \pm 0 $ & $ 0 \pm 0 $ & $ 3 \pm 1 $ & $ 3 \pm 1 $ &\cite{wei09} \\
Observed & -- & -- & 0 & 0 & 2(?) & 2 & \\
\hline
\end{tabular}

\caption{Simulated post-AGB population for various models in the
    region of the Saffer et al. (1997) complete sample.}
$\alpha$ indicates an $\alpha$-enhanced initial composition.

\label{tab:pagb_list}
\end{minipage}
\end{table}
The resulting post-AGB numbers differ greatly from one track to
another and there is a general trend with mass and
metallicity. Fig.\,\ref{fig:plots_0546} shows a simulated post-AGB
population, within the brightness and positional criteria set out,
assuming a mass of 0.546\msun. We run the simulation for each post-AGB
evolutionary track we have obtained and summarise the most relevant
tracks in Tab.\,\ref{tab:pagb_list}. The masses stated are the final
post-AGB/WD and the initial ZAMS in their respective papers. The
metallicities are the initial compositions of the stars on the MS. The
numbers for each population and the total are given. The reference for
the model used is stated in the final column. The original simulation
contains a multiple of the evolved stars present in our galaxy. A
normalisation factor is calculated and a random set of stars selected
for the multi-Galaxy. This way 100 synthetic representations of the
Galaxy are produced.

\section{Conclusion} 
The observed PG subsample of \cite{saf97} implies that there are very
few post-AGBs in the halo and the ones that exist have low masses ($M
< 0.55$\msun). The masses are very much in agreement with halo WD mass
distributions both within the Milky Way and other
galaxies \cite{pau06,lie05}. However, our synthetic galactic model
shows that the lower the mass of the central star (and progenitor) the
slower the evolution and the number of stars meeting our brightness
and temperature criteria will increase. This is displayed in Fig.\,16
of \cite{wei09}. In the same figure, a small metallicity effect can be
seen and this is reflected in our numbers but this is fairly small
effect at sub-solar metallicities. Our results suggest, as the
observational fields are complete, the evolutionary paths or
timescales for the majority of halo stars differ from the
theory. Increasing the central star evolutionary speed across the HR
diagram would bring the observed and theoretical populations in
agreement, however, a PN would be a likely result. Even fewer PN are
known in the halo than post-AGBs ruling out that option. An
alternative solution is that the majority of evolved stars in the halo
do not ascend the AGB. Obviously, this would reduce the expected
number of post-AGB stars and would also be consistent with the HB
post-AGB ratio observed in the \cite{saf97} sample. A similar HB to
post-AGB ratio is observed in M32 by \cite{bro08}. \citet{bro08}
propose that this is unlikely due to an increase in evolutionary speed
or circumstellar absorption. The post-AGB population may not be
observed as they do not exist. Instead of ascending the AGB, the
pop. II halo stars would evolve via the EHB and straight on to the WD
cooling track. If this is the case for all such populations then there
would be implications for subsequent galactic evolution.


\begin{theacknowledgments}
We would like to thank Achim Weiss and Peter Wood for their help and
co-operation for their respective post-AGB tracks. S.W.  would like to
thank the Science and Technology Facilties Council (STFC) and the
organising committee for their financial support. S.C. acknowledges
financial support from the European Commission in the form of a Marie
Curie Intra European Fellowship.
\end{theacknowledgments}



\bibliographystyle{aipproc}   

\bibliography{sweston}

\end{document}